# Operational Experience of the NML Cryogenic Plant at the FAST Test Facility

**Timothy Wallace, Joaquim Creus-Prats, Joseph Hurd, Michael J White, Jerry Makara, Liujin Pei, Benjamin Hansen, Jay Theilacker, Rick Bossert, Alexander Martinez, James K Santucci, Sasha Romanov**

Fermi National Accelerator Laboratory, PO Box 500, Batavia, IL 60510

twallace@fnal.gov

**Abstract.** The NML cryogenic plant cools two individually cryostated superconducting radio frequency (SRF) capture cavities and one prototype ILC cryomodule with eight SRF cavities. This complex accelerates electrons at 150 MeV for the Integrable Optics Test Accelerator (IOTA) ring, located at the Fermilab Accelerator Science and Technology (FAST) facility. The cryogenic plant is composed of two nitrogen precooled Tevatron satellite refrigerators, two Mycom 2016C compressors, a cryogenic distribution system, a Frick purifier compressor, two charcoal bed adsorber purifiers, and a liquid ring vacuum pump with a roots booster. The SRF cavities are immersed in a 2.0 K liquid helium bath, shielded with a 5 K gaseous helium shield and a liquid nitrogen cooled thermal shield. Since 2019, this R&D accelerator complex has gone through four science runs with an average duration of 12 months. Operational experience for each run, availability metrics, performance data and common outages are presented in this paper.

## 1. Introduction

The New Muon Laboratory (NML) cryogenic plant provides cryogens to the two individually cryostated superconducting radio frequency (SRF) capture cavities (CC1 & CC2) and one prototype ILC cryomodule with eight SRF cavities. The cryogenic plant consists of two nitrogen precooled Tevatron satellite refrigerators in parallel. Each satellite refrigerator includes a set of wet and dry reciprocating expansion engines, a STAR heat exchanger, and a 500 L helium dewar. Testing on Tevatron satellite refrigerators has shown that each have a capacity of 3.7 g/s liquefaction and 50 W at 4.5K [1]. The expansion engines are completely overhauled with the pistons, flywheels, and all seals replaced after at least 8,000 hours of runtime or annually to prevent failures during operations [2].

The refrigerators are supplied with up to 120 g/s of 20 barg helium from two model 2016C Mycom screw compressors that operate in parallel. A Frick purifier compressor, also in parallel to the Mycoms, operates at flow rates up to 13 g/s. The supply and discharge of the compressors are connected to a 113,562 Liter (30,000-gallon) gaseous helium storage tank with inventory control valves that regulate suction and discharge to nominal operating pressures. Before entering the refrigerators, a portion of the helium stream is purified via two LN2 cooled absorbers that are each capable of handling up to 60 g/s or the maximum flow from one Mycom compressor. Each absorber can operate in one of three modes. In full flow mode the absorber is aligned with Mycom discharge. In LN2 economy mode the absorber is aligned with the Frick compressor. This mode limits the amount of LN2 boiled off with the reduced helium flow from the Frick. Lastly, the absorber must periodically be placed in regeneration mode where the absorber is temporarily isolated and warm nitrogen gas is pushed through the charcoal media. This typically occurs at startup and as needed during operation as the media becomes saturated with impurities. Ideally, the absorbers are set in LN2 economy mode where the Frick is inline with the absorbers. This reduces LN2 consumption and allows one absorber to be regenerated at a time without affecting operations. The Frick compressor is also directly downstream of the sub-atmospheric portion of the circuit. In LN2 economy mode, any contamination from air ingress is immediately diverted through the purifiers.

The refrigerators are connected to the cryomodule and capture cavities via a transfer line that houses a 5K helium supply, LN2 supply, and 4.5K helium return. The LN2 supply is used to cool the 80K thermal shields of the transfer line, cryomodule, and capture cavities. The cryomodule and capture cavities each have a JT heat exchanger and JT valve in series fed by the 5K helium supply used for filling the cryostats. Upstream of each JT heat exchanger is a bypass that cools the 5K thermal shield of each cryostat before flowing back to the cryoplant through the transfer line 4.5K helium return. In addition, upstream of the cryomodule is a mixing chamber that can mix 5K supply with ambient temperature helium for warmup and cooldown.

2K operation within the cryomodule and capture cavities is achieved with a Kinney Vacuum Skid (KVS). The KVS consists of a Kinney model KMBD 10,000 roots blower and a Kinney model KLRC 2100 liquid ring pump with capacity to pump down the cryomodule and capture cavities together or individually to 3066 Pa via a long pumping header. The pumping header and KVS were designed for sub-atmospheric operation though not with helium guard. As such, if leaks were to develop at seals, air ingress or contamination is possible during 2K operations.

The cryogenic plant was originally constructed and commissioned in the early 2010s [1]. Since 2019, NML has gone through four science runs with an average duration of 12 months. Operational experience for each run, availability metrics, performance data, and common outages are presented below.

**2. NML 2K Availability**

Availability of the NML cryogenic plant throughout the four science runs was tracked by categorizing each day that the cryogenic system was operational into one of the following: Cryo Ready, SRF Hours, Utility Downtime, and Cryo Downtime. Any day where the cryomodule is powered with SRF is accounted as a full day (24 hours) of SRF Hours. In reality SRF mainly operates with onsite operators during the daytime. Cryo Ready represents periods of time when the cryomodule is at 2K operation and SRF is not powered. Any day that the cryomodule is not at 2K while the cryogenic system is operational is then categorized into either Utility Downtime or Cryo Downtime, depending on the reason the cryomodule is not at 2K. Cryo Downtime includes cooldown and warmup of the cryogenic plant, as well as any cryogenic system originated interruptions that prevent it from operating at 2K. Interruptions related to the cryogenic support systems (power outages, cooling water, instrument air, etc.) are labelled as Utility Downtime. For simplicity, days where the cryomodule is mostly at 2K (greater than half the day) are considered at 2K for the entire day. With the categories defined, the monthly availability of the NML cryogenic system during the four science runs is presented in Figures 1 through 4. Following each figure, is a brief discussion of the major interruptions in 2K availability.

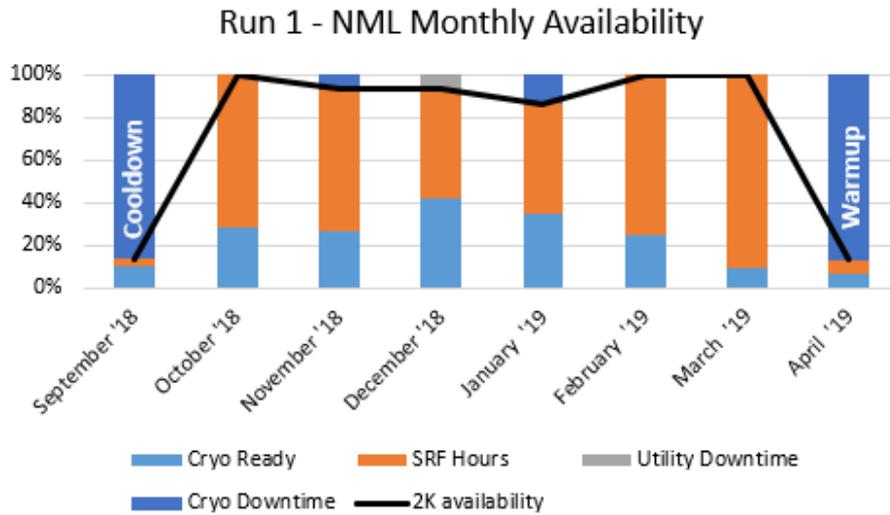

**Figure 1.** Run 1 monthly availability. Run 1 was from September 2018 to April 2019.

Run 1 started in september 2018 and lasted approximately 8 months, with system remaining at 2K operation throughout most the entire duration. The Kinney Vacuum Skid (KVS) tripped once in December due to a cooling water pipe burst (utility downtime) and once in January due to a Mycom electrical fault (cryo downtime) that tripped the Mycom compressor first and then the KVS after a rise of suction pressure. These were the only two significant sources of downtime for the system.

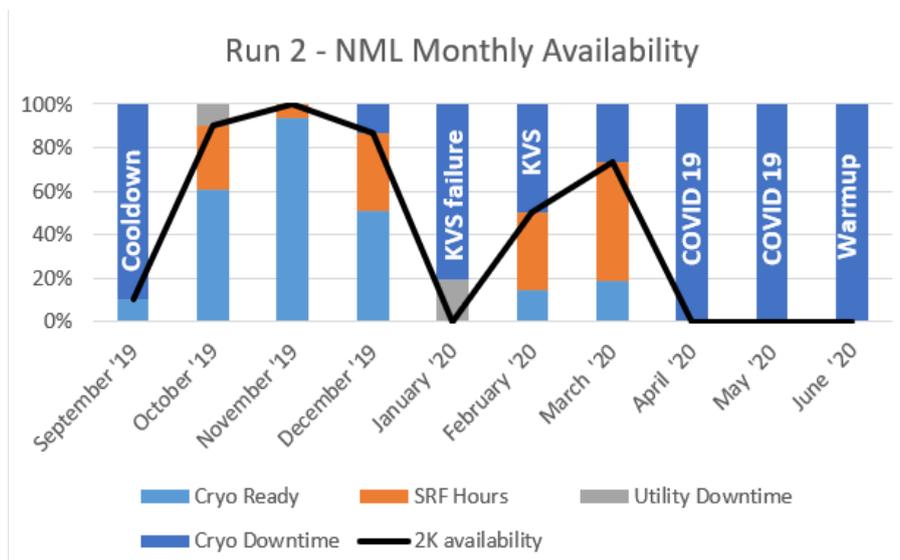

**Figure 2.** Run 2 monthly availability. Run 2 was from September 2019 to June 2020.

Run 2 started in September 2019 and lasted 10 months. It was severely impacted by a KVS booster pump failure and was ultimately stopped short due to the impact from the COVID-19 pandemic. The booster pump failed in January 2020 and ultimately took 6 weeks to repair. A spare pump was available at the time of the failure but the cause was determined to be in part due to excessive load on the pump bearings. As such, repair included not only replacement of the booster pump but relocation of the booster motor axially closer to the pump to reduce bearing load from the belt. A new motor base plate was fabricated and installed and the pump shaft seal was modified to accomodate the closer belt. Completion of the repair necesitated placement of a rigging contract to replace the pump and relocate the motor. Procurement of the rigging contract and scheduling of the work was a major contributor to the long downtime and highlights the importance of maintaing the capability of installing spare components either inhouse or with open contracts.

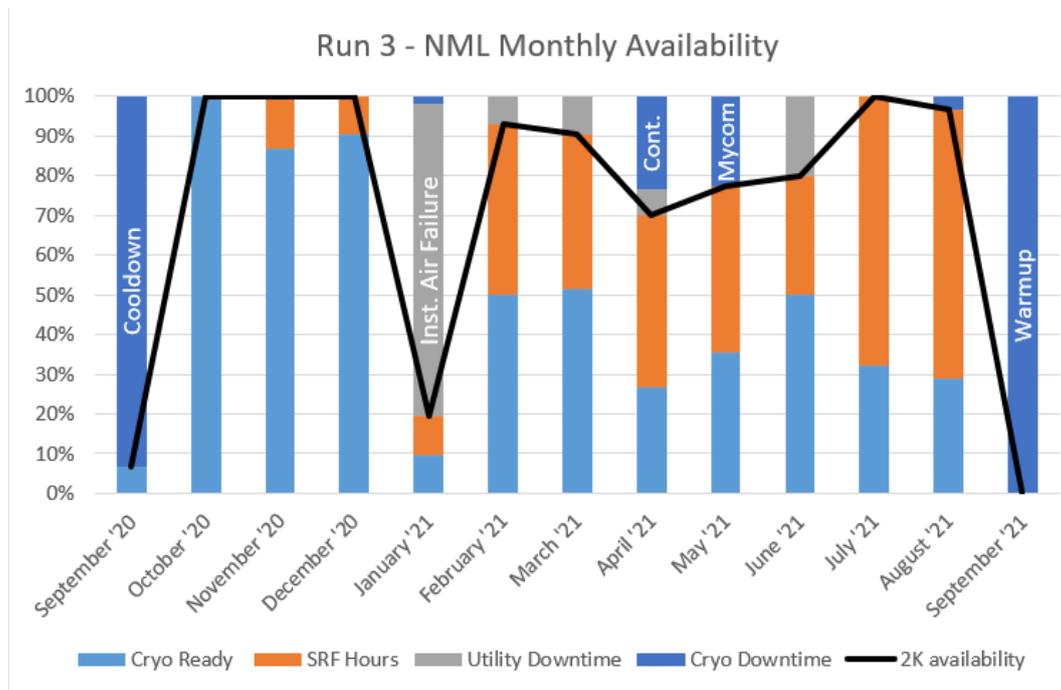

**Figure 3.** Run 3 monthly availability. Run 3 was from September 2020 to June 2020.

Run 3 started in September 2020 and extended to over one full year. Four months into the run, the primary instrument air compressor failed. The secondary air compressor was immediately brought online and promptly failed within hours. Both failures were related to the compressors electronic boards. Vaporized nitrogen from the LN2 dewar was then brought inline as instrument air backup but could not meet demand and thus operability of the cryogenic system was limited. The system could not achieve 2K operations until a new air compressor with sufficient capacity was brought online after 4 weeks of downtime.

In April 2021, after approximately 4,200 hours of 2K operation (nearly the total amount of hours at 2K for run 1 and more than run 2), contamination issues resulted in cryomodule JT valves clogged to the point where nominal liquid helium levels within the cryomodule and capture cavities could not be maintained. Localized thermal bumps were performed to flush condensed contamination through to the absorbers. Thermal bumps included isolating flow in suspected lines for long enough periods to allow the temperature to increase to ~100K. Depending on the location of the thermal bump, warm helium supply could also be injected to aid in warmup. The thermal bumps were successful but necessitated over a week of downtime.

In May 2021, two separate failures amounted to a little over a week of downtime. One of the expansion engine flywheels shafts mechanically failed and one of the Mycom's tripped due to what was believed to be a PLC software issue. It's noted that the flywheel shaft failure occured at only 5,500 hours and thus within the recommended overhaul period of 8,000 hours.

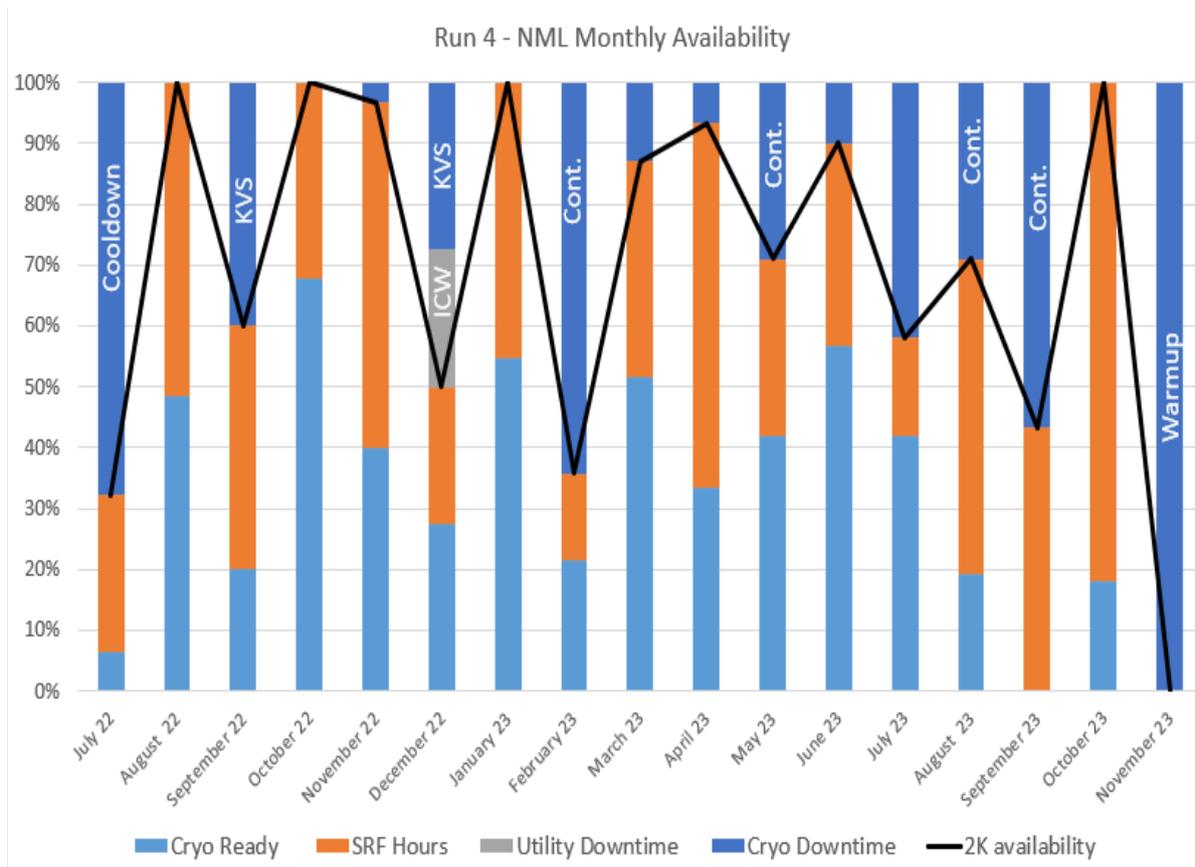

**Figure 4.** Run 4 monthly availability. Run 4 was from July 2022 to November 2023.

Run 4 was the longest run starting in July 2022 and lasting until warmup of the system in November 2023, totalling 17 months. A number of issues throughout the run were dealt with that interrupted 2K operation. The two most prevalent issues that amounted to the greatest downtime were problems related to contamination (abreviated cont. in Figure 4) and repeated KVS trips. Early during the run the Frick purifier compressor failed with a suspected helium to water leak in the heat exchanger. A replacement heat exchanger was not available at the time of the failure and so the Frick was not operable during the entire run. The helium/water leak may have contributed to the contamination issues faced during this run. Additionally, without the Frick compressor, the Mycom compressor discharge flow was directed through the absorbers. As a result, anytime the absorbers were regenerated, which occured 2 times each during run 4, the capacity of the plant was greatly reduced and 2K operations could not be achieved.

Contamination issues were first observed in February 2023 after approximately 4,000 hours of 2K operation. The cryomodule and transfer line were brought to 100K in an attempt to flush any contamination through the system. The South refrigerator was also isolated and U-tubes suspected of being clogged were pulled for inspection. No obvious blockage was observed during the inspections but spikes in contamination were measured as the system was flushed. In total, the thermal bumps with inspections resulted in 20 days of downtime.

In May 2023, differential pressure accross the absorbers rose to the point where regeneration was necessary. The differential was likely due to air contamination freezing in the absorber heat exchangers. It was also noticed that one of the Mycom compressor discharge relief valves was popping open below the relief set pressure relieving helium back to compressor suction. Both compressor outlets flow completely through the absorber and thus higher absorber differential pressure requires higher compressor discharge pressure to achieve the same plant capacity. The high compressor discharge contributed to the relief valve opening prematurely. Regeneration of the absorbers and replacement of the relief valve resulted in another 10 days of downtime.

Two additional thermal bumps were performed in August and September with the absorbers again regenerated during the second of the two thermal bumps. It was clear that a more thorough flushing of the system was necessary but it was believed that the science run was near completion and thus shorter flushes would be sufficient.

The KVS tripped multiple times throughout run 4. The function of the KVS is to pumpdown and maintain pressure in the cryomodule and capture cavities at 3 kPa. When the KVS trips, and isn't brought back online fast enough, pressure rises to compressor suction, ~108 kPa, bringing the cryomodule up to saturation temperature or 4.1K. Once the KVS is brought back online, the cryomodule then needs to be pumped back down which takes 3-4 hours. This means that even simple trips can result in a lost day of SRF operation. The KVS trips were caused by a multitude of reasons. One persistent issue was a noisy pressure signal where single pressure readings were spiking above the high pressure trip setpoint. The high pressure readings were eventually determined not to be real and thus the trip was altered to only occur when the average of many readings exceeded the setpoint. Other causes of KVS trips included cooling water leaks developing and electrical equipment failures. The repeated nature of the trips highlights the importance of protecting against single point failures.

### 3. Performance Metrics

Two performance metrics are proposed herein to assess the availability of the NML cryogenic system during the four science runs. The first metric, titled Cryo Availability, assesses the availability of the cryogenic system to be at 2K and is defined as follows:

$$Cryo\ Availability = 100 * \frac{downtime\ weeks}{downtime\ contingency} \qquad (1)$$

The cumulative downtime weeks are summarized in Figures 1 through 4 for each science run. Downtime contingency was estimated based off the expected duration of a run. Before the fourth and last science run, APS-TD Cryogenics and AD-FAST agreed to target six months of SRF active operation per year to estimate costs associated with cryogenic operation and LN2 consumption. To achieve six months of SRF active operation, the cryogenic system must be brought online, maintained, and safely shut down. A total of 12 weeks of downtime contingency is estimated where the cryogenic system is unavailable broken out into the four durations listed below. This total was based off operational experience at NML and similar cryogenic plants at other Fermilab test sites.

- 4 weeks for cool-down
- 4 weeks for warm-up
- 2 weeks of Utility downtime

- 2 weeks of Cryo downtime

The second performance metric is titled 2K Usage. This assesses utilization of the cryogenic system when at 2K operation and is defined as follows:

$$2K\ Usage = 100 * \frac{SRF\ Hours + Cryo\ Ready}{SRF\ 2K\ requested + SRF\ contingency} \tag{2}$$

SRF Hours and Cryo Ready are defined in Section 2 and the SRF 2K requested is 6 months as previously discussed. An SRF contingency of 2 weeks is estimated based off operational experience at NML. Table 1 presents the calculated metrics for the four science runs. A target of less than 100% is proposed for both metrics. Anything less than 100% for Cryo Availability indicates that the downtime during the run was less than what was budgeted for in contingency. A 2K Usage less than 100% indicates that the total time the cryomodule and capture cavities were powered with SRF or ready is less than the requested time by AD-FAST with contingency.

**Table 1**: Performance Metrics for the Four Science Runs

|  | Target | Run 1 | Run 2 | Run 3 | Run 4 |
|---|---|---|---|---|---|
| Cryo Availability | <100% | 69% | 209% | 113% | 149% |
| 2K Usage | <100% | 101% | 70% | 156% | 172% |

## 4. Conclusions

The four science runs were completed successfully with over 22,000 hours of 2K operation. Careful execution of standard operating procedures (including scheduled maintenance) along with prompt alarm response and equipment repairs (sometimes off-hours) allowed for limited interruptions. NML is a complex cryogenic plant. Failures in legacy rotary equipment and operations falling outside nominal are sometimes unavoidable. With that in mind, certain lessons were reinforced throughout the four runs.

Contamination issues occured in both runs 3 and 4 after 4,000 hours of 2K operation. As previously noted, a portion of the cryogenic plant operates subatmospheric when the system is at 2K. The potential for contamination buildup increases when subatmospheric as air ingress can occur. After each run, attempts were made to identify potential leak sources but no obvious sources were found. It's possible that the contamination built up in the system slowly over multiple small leaks. Thermal bumps were successful in bringing the system back to 2K operation for short durations, but issues with contamination persisted as each run continued. 4,000 hours equates to just under 6 months, the target for 2K active operation. It seems reasonable that a complete system warmup should be preplanned after 6 months of 2K operation. This would allow a more complete flushing of the system as well as an opportunity to perform any system maintenance.

Managing single point failures and maintaining hot spares is crucial. The repeated KVS trips show the risk of being susceptible to single point failures. Building redundancy in design is ideal though cost and schedule are additional considerations. Ideally, in place of the NML KVS would be 3 smaller vacuum pump skids where only 2 are needed to operate at 2K.  This is not always possible and thus emphasis must be placed on ensuring reliability.  This includes keeping up with maintenance, tracking common failures, and maintaining hot spares.  It also includes possessing the ability to install the spares should a failure occur.  This was an issue which caused downtime during run 2 when the KVS booster pump failed.

Lastly, it's important to ensure operability of backup or spare components.  The backup air compressor failing hours after the primary failed resulted in a 4 week downtime during run 3.  A best practice is to periodically energize or put in service backup equipment to prove operability.  This lesson can also be serve as a reminder to perform routine maintenance on hot spares in addition to installed equipment.

**Acknowledgments**
This work was produced by FermiForward Discovery Group, LLC under Contract No. 89243024CSC000002 with the U.S. Department of Energy, Office of Science, Office of High Energy Physics. Publisher acknowledges the U.S. Government license to provide public access under the DOE Public Access Plan DOE Public Access Plan.